\pdfoutput=1
\documentclass[fleqn,usenatbib]{mnras}

\usepackage{newtxtext,newtxmath}

\usepackage[T1]{fontenc}
\usepackage{ae,aecompl}

\usepackage[utf8]{inputenc}

\usepackage{graphicx}	
\usepackage{amsmath}	
\usepackage{bm}         
\usepackage{amssymb}	

\usepackage[usenames, dvipsnames]{color}

\newcommand{\Lagr}{\mathcal{L}}
\graphicspath{{figures/}{figures/tess-sim/}{figures/diamonds/}}

\title[Gaussian Processes applied to red-giant stars]{Gaussian Process modelling of granulation and oscillations in red-giant stars}

\author[Filipe Pereira et al.]{
Filipe Pereira$^{1,2}$\thanks{Email: Filipe.Pereira@astro.up.pt}, Tiago L.~Campante$^{1,2,3}$, Margarida S.~Cunha$^{1,2}$, Jo\~{a}o P.~Faria$^{1}$,
\newauthor Nuno C.~Santos$^{1,2}$, Susana C.~C.~Barros$^{1,2}$, Olivier Demangeon$^{1}$, James S.~Kuszlewicz$^{4,5}$
\newauthor and Enrico Corsaro$^{6}$
\\
$^{1}$Instituto de Astrofísica e Ciências do Espaço, Universidade do Porto, CAUP, Rua das Estrelas, PT4150-762 Porto, Portugal\\
$^{2}$Departamento de Física e Astronomia, Faculdade de Ciências, Universidade do Porto, Rua do Campo Alegre, PT4169-007 Porto, Portugal\\
$^{3}$Kavli Institute for Theoretical Physics, University of California, Santa Barbara, CA 93106-4030, USA \\
$^{4}$Max Planck Institute for Solar System Research, D-37077 G\"ottingen, Germany \\
$^{5}$Stellar Astrophysics Centre (SAC), Department of Physics and Astronomy, Aarhus University, Ny Munkegade 120, DK-8000 Aarhus C, Denmark \\
$^{6}$INAF – Osservatorio Astrofisico di Catania, via S. Sofia 78, 95123 Catania, Italy
}

\date{Accepted XXX. Received YYY; in original form ZZZ}

\pubyear{2019}

\begin{document}
\label{firstpage}
\pagerange{\pageref{firstpage}--\pageref{lastpage}}
\maketitle

\begin{abstract}
	The analysis of photometric time series in the context of transiting planet surveys suffers from the presence of stellar signals, often dubbed ``stellar noise''. These signals, caused by stellar oscillations and granulation, can usually be disregarded for main-sequence stars, as the stellar contributions average out when phase-folding the light curve. For evolved stars, however, the amplitudes of such signals are larger and the timescales similar to the transit duration of short-period planets, requiring that they be modeled alongside the transit. With the promise of TESS delivering on the order of $\sim\!10^5$ light curves for stars along the red-giant branch, there is a need for a method capable of describing the ``stellar noise'' while simultaneously modelling an exoplanet's transit. In this work, a Gaussian Process regression framework is used to model stellar light curves and the method validated by applying it to TESS-like artificial data. Furthermore, the method is used to characterize the stellar oscillations and granulation of a sample of well-studied \textit{Kepler} low-luminosity red-giant branch stars. The parameters determined are compared to equivalent ones obtained by modelling the power spectrum of the light curve. Results show that the method presented is capable of describing the stellar signals in the time domain and can also return an accurate and precise measurement of $\nu_\text{max}$, i.e., the frequency of maximum oscillation amplitude. Preliminary results show that using the method in transit modelling improves the precision and accuracy of the ratio between the planetary and stellar radius, $R_p/R_\star$. The method's implementation is publicly available\footnotemark.
\end{abstract}




\footnotetext{\url{https://github.com/Fill4/gptransits}}
\section{Introduction}\label{sec:intro}
Red-giant stars show significant stellar signals on timescales ranging from hours to weeks \citep[e.g.,][]{mathur_2011,kallinger_2014,kallinger_2016,north_2017}. In particular, granulation is responsible for introducing correlated noise in the time series photometry of red giants, with a characteristic timescale similar to the transit duration of a planet in a close orbit. Such ``stellar noise'' should therefore be taken into account in the modelling of planetary transits if biases on the fitted transit parameters are to be avoided \citep{carter_2009,barclay_2015}. Moreover, solar-like oscillations characterized by large mode amplitudes are also present in the time series photometry of red giants and ultimately require similar treatment \citep{grunblatt_2016,grunblatt_2017}.

With NASA's TESS mission \citep{ricker_2015,huang_2018,vanderspek_2018} expected to deliver on the order of $\sim\!10^5$ light curves (from full-frame images) for stars along the red-giant branch \citep{sullivan_2015,campante_2016b,campante_2018}, a need becomes apparent for a method capable of describing such stellar signals while simultaneously modelling an exoplanet's transit. Since stellar granulation is a stochastic phenomenon, there is no functional form capable of describing it in the time domain. Nonetheless, there are tools that, while not leading to a parametric model, are capable of defining a model based on a set of properties of the data. One such method consists in the use of Gaussian Process (GP) regression \citep{rasmussen_2006}. This method has been increasingly adopted in a variety of scenarios, including previous modelling of light curves \citep[e.g.,][and references therein]{barclay_2015,grunblatt_2016,grunblatt_2017} and radial-velocity time series \citep{brewer_2009,haywood_2014,faria_2016a,farr_2018}.

When employing this framework to the modelling of stellar signals in photometric time series it is common, however, to pay little attention to the physical meaning of the resulting model parameters. In the aforementioned works, the main focus has been the convergence of the adopted model whilst the values obtained are rarely analysed to assess their physical validity with respect to the type of star being observed. A main goal of this work is therefore to evaluate whether it is possible to recover physically meaningful parameters when using GP regression to model stellar signals in the time domain.

The paper starts with an introduction to the theoretical framework of GP regression in Sect.~\ref{sec:GPs}. In order to assess the validity and accuracy of the method, the GP regression is first applied to artificial time series similar to those expected from the TESS mission (Sect.~\ref{sec:appTESS}). The parameters obtained in such fits are then compared to the true values injected in the simulated light curves. Subsequently, the method is tested with real \textit{Kepler} data for low-luminosity red-giant branch (LLRGB) stars (Sect.~\ref{sec:appKepler}). Here, the aim is to compare the parameters obtained using GP regression to the equivalent parameters obtained when performing a standard fit to the power spectrum of the time series. Conclusions are drawn in Sect.~\ref{sec:conclusions}.

\section{Gaussian Process regression}\label{sec:GPs}

\subsection{Theoretical framework}

Gaussian Processes are non-parametric models capable of describing correlated stochastic signals. They describe each point as a correlated random variable with a mean value and a variance, where any finite collection of those variables has a multivariate Gaussian distribution. The measure of similarity, or correlation, between pairs of points in the signal with respect to the distance between them in time is given by the covariance function or kernel of the GP. The covariance matrix, $\textbf{K}$, is defined as:
\begin{equation}
	K_{ij} = \sigma_i^2 \delta_{ij} + k_{\bm\alpha}(\tau_{ij}) \, ,
	\label{eq_covariance}
\end{equation}
where for each pair of points $i$ and $j$, $\sigma_i$ is the uncertainty of the observation, $\delta_{ij}$ is the Kronecker delta, and $k_{\bm\alpha}(\tau_{ij})$ is the kernel, with $\tau_{ij} = |t_i - t_j|$ being the absolute distance in time between points $i$ and $j$. As indicated by the notation, the kernel depends on a set of parameters, $\bm\alpha$.

GP regression consists in the selection of a kernel that describes the correlation between data points and then finding the set of parameters that best represent the observed data. The merit of the fit can be determined using the log-likelihood function
\begin{equation}
	\textnormal{ln}\Lagr(\textbf{r}) = -\frac{1}{2} \textbf{r}^\text{T} \textbf{K}^{-1} \textbf{r} - \frac{1}{2} \textnormal{ln} \vert \textbf{K} \vert  - \frac{n}{2}\textnormal{ln}(2\pi) \, ,
	\label{eq_likelihood}
\end{equation}
where $\textbf{r}$ are the residuals after removing the mean model from the data (this mean model may represent any parametric model used to characterize the data, e.g., a transit model) and $n$ is the number of data points.

When using GP models, the most important decision to be made concerns the selection of an appropriate kernel. A number of aspects have to be taken into account when performing such a selection. Standard computation of Eq.~(\ref{eq_likelihood}) has complexity of order O($n^3$), meaning that computation time increases very rapidly with the number of observations. Efforts have been made to reduce this complexity \citep{ambikasaran_2015,foreman-mackey_2017} and hence computational time. However, these efforts inevitably carry with them a number of shortcomings, especially with respect to the functional form of the kernels permitted.

\subsection{Algorithm}
\label{sec:algorithm}

In this work, the implementation of GP regression chosen was the {\sc Python} package {\sc celerite}\footnote{\url{https://github.com/dfm/celerite}} \citep{foreman-mackey_2017}. This package applies some restrictions to the functional form of the kernels in order to achieve a computational complexity of order O($n$), an acceptable level if one desires to perform fast computations on long time series, like the ones provided by TESS.

The {\sc celerite} implementation requires that the kernel $k_{\bm\alpha}$ be a mixture of exponential functions:
\begin{equation}
	k_{\bm\alpha} (\tau_{ij}) = \sum^M_{m=1} a_m \ \text{exp} (-c_m \ \tau_{ij}) \, ,
	\label{eq_sum_exp}
\end{equation}
where $\bm\alpha = \{a_m, \ c_m\}$. By introducing complex parameters, $a_m \rightarrow a_m \pm i \ b_m$ and $c_m \rightarrow c_m \pm i \ d_m$, and rewriting the exponentials in Eq.~(\ref{eq_sum_exp}) as sums of sines and cosines, the equation can be written as a mixture of quasi-periodic oscillators:
\begin{equation}
	\begin{split}
		k_{\bm\alpha} (\tau_{ij}) = \sum^M_{m=1} \ [ & a_m \ \text{exp} (-c_m \ \tau_{ij}) \ \text{cos} (d_m \ \tau_{ij}) \\
		&+ b_m \ \text{exp} (-c_m \ \tau_{ij}) \ \text{sin} (d_m \ \tau_{ij}) \ ] \, ,
	\end{split}
	\label{eq_sum_sines}
\end{equation}
this time with $\bm\alpha = \{a_m, \ b_m, \ c_m, \ d_m\}$.

To provide physical insight into the previous equation, consider the power spectral density (PSD) of a stochastically-driven, damped harmonic oscillator:
\begin{equation}
	S( \omega ) = \sqrt{\frac{2}{\pi}} \frac{S_0\omega_0^4 }{ ( \omega^2 - \omega_0^2)^2 + \omega_0^2 \omega^2 / Q^2} \, ,
	\label{eq_psd_sho}
\end{equation}
where $\omega$ is an angular frequency, $\omega_0$ is the frequency of the undamped oscillator, $Q$ is the oscillator's quality factor, and $S_0$ is proportional to the power of the spectrum at $\omega$ = $\omega_0$, i.e.,
\begin{equation}
	S(\omega_0) = \sqrt{\frac{2}{\pi}} \ S_0 \ Q^2 .
\end{equation}
Following \citet{foreman-mackey_2017}, the PSD in Eq.~(\ref{eq_psd_sho}) can be used to describe the power spectrum of Eq.~(\ref{eq_sum_sines}), and a relation can thus be written between $\bm\alpha$ and the parameters $S_0, \ Q \ \text{and} \ \omega_0$. Doing so, the kernel in Eq.~(\ref{eq_sum_sines}) translates to
\begin{equation}
	k(\tau; \ S_0, \ Q, \ \omega_0) = S_0 \ \omega_0 \ Q \ \text{e}^{\frac{\omega_0\tau}{2Q}} \: \times \nonumber\\[2pt]
\end{equation}
\vspace{-10pt}
\begin{align}
	\begin{cases}\!
		\textnormal{cosh}(\eta \omega_0 \tau) + \frac{1}{2 \eta Q} \ \textnormal{sinh}(\eta \ \omega_0 \ \tau) \, , & 0 < Q < 1/2 \, , \\[2pt]
		2(1 + \omega_0 \tau) \, ,                                                                                   & Q = 1/2 \, ,     \\[2pt]
		\textnormal{cos}(\eta \omega_0 \tau) + \frac{1}{2 \eta Q} \ \textnormal{sin}(\eta \ \omega_0 \ \tau) \, ,   & 1/2 < Q \, ,     \\[2pt]
	\end{cases}
	\label{eq_sho}
\end{align}
where $\eta = \vert 1- (4Q^2)^{-1}\vert^{1/2}$.

The work of \citet{foreman-mackey_2017} remarks upon some limits of physical interest regarding the power spectrum in Eq.~(\ref{eq_psd_sho}) which translate also to the kernel in Eq.~(\ref{eq_sho}). When considering the limit $Q = 1/\sqrt{2}$, the kernel simplifies to
\begin{equation}
	k(\tau) = S_0 \ \omega_0 \ \text{e}^{-\frac{1}{\sqrt{2}} \omega_0 \tau} \text{cos} \left( \frac{\omega_0 \tau}{\sqrt{2}} - \frac{\pi}{4} \right) \, ,
	\label{eq_kernel_gran}
\end{equation}
with the corresponding PSD becoming
\begin{equation}
	S(\omega) = \sqrt{\frac{2}{\pi}} \frac{S_0}{(\omega/\omega_0)^4+1} \, .
	\label{eq_psd_gran}
\end{equation}
Equation (\ref{eq_psd_gran}) has the same functional form as the equation commonly used to model the granulation in a power spectrum analysis \citep[e.g.,][]{kallinger_2014}. The exponent in the denominator was originally set to 2 by \citet{harvey_1985} to model the solar background signal, with an exponent of 4 found to be more appropriate by \citet{kallinger_2010} for red giants. Since Eq.~(\ref{eq_psd_gran}) corresponds to the PSD of the kernel in Eq.~(\ref{eq_kernel_gran}), this kernel can be used to capture the same granulation signal in the time domain.

The other interesting limit of Eq.~(\ref{eq_sho}) is that, for values of $Q>1$, the PSD of the kernel can be used as an approximation to the Gaussian-like shape of the oscillation bump (or power excess) found in the power spectrum, meaning that this kernel can be used to capture the signal from this oscillation bump in the time domain.

\begin{figure}
	\centering
	\includegraphics[width=\columnwidth]{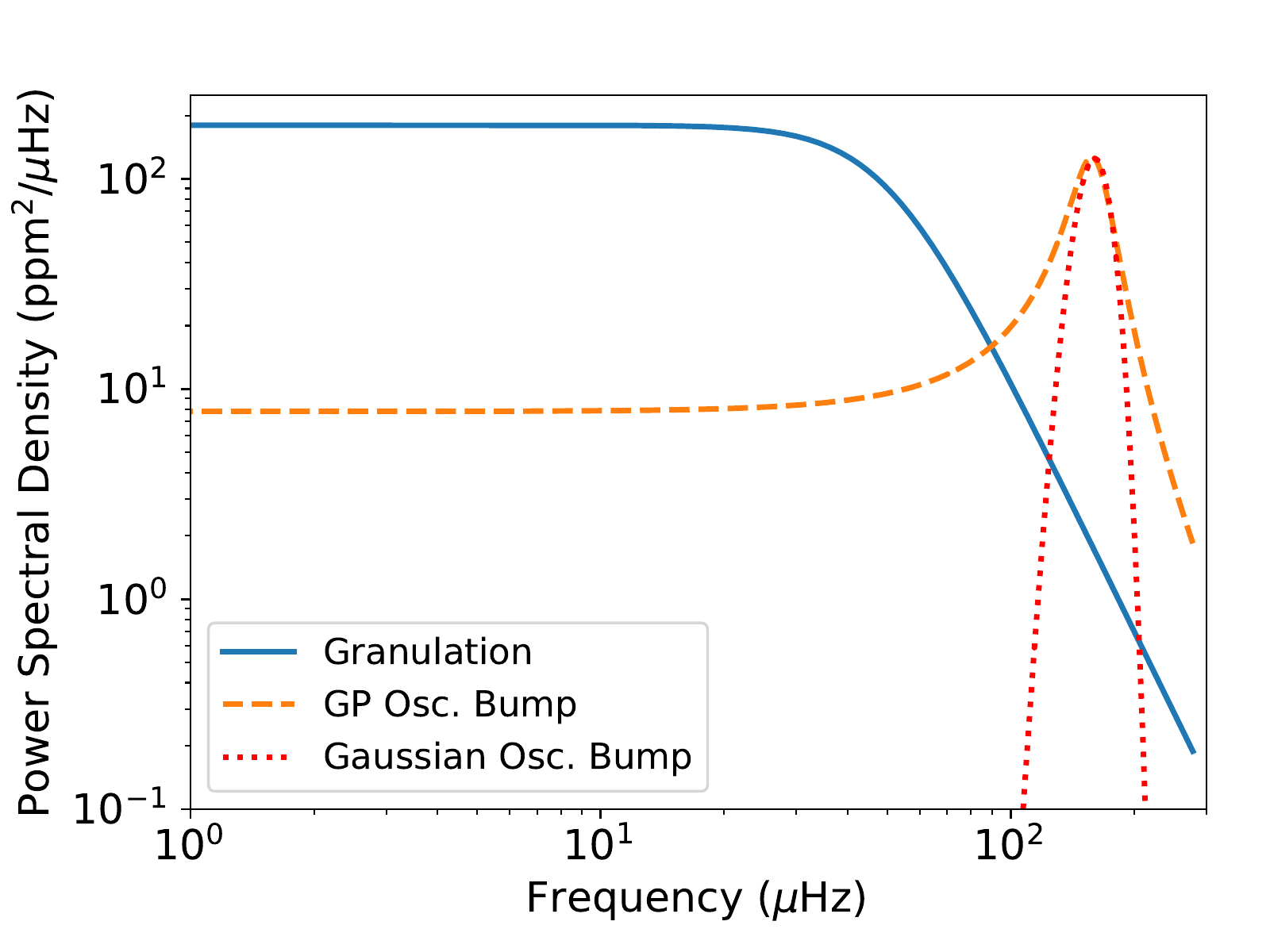}
	\caption{Solid (blue) curve depicts the power spectral density of a granulation profile (Eq.~\ref{eq_psd_gran}). Dashed (orange) and dotted (red) curves are the power spectral densities of the functions used to capture the signal from the oscillation bump in a GP and power spectrum analyses, respectively.}
	\label{fig:comparison_kernel_psd}
\end{figure}

\subsection{Implementation}
\label{sec:implement}

The two models adopted in this work are a sum of the kernels described above. The first model, Model 1, contains a single granulation kernel (Eq.~\ref{eq_kernel_gran}) to capture the mesogranulation, an oscillation bump kernel (Eq.~\ref{eq_sho}; $1.2 < Q < 18$, where the limits were defined empirically for this work), as well as a white noise kernel (which is simply a constant value added in quadrature to the diagonal of the covariance matrix). The second model, Model 2, results from adding a second granulation kernel to Model 1. This stems from the fact that, as \citet{kallinger_2014} observed, at least the granulation and mesogranulation components should be adopted when estimating the stellar background signal for red-giant stars, owing to granulation being relevant at different timescales. In the literature a third component is sometimes also added to model long-term variations \citep{kallinger_2014,corsaro_2015b} (on the order of 1 cycle per 11 days). However, since typical TESS time series are considered in this work (i.e., with a duration of 27.4 days), such long-term variations would not be detectable and hence a third component is discarded.

Study of stellar signals in the literature is commonly performed in the frequency domain by means of power-spectrum fitting. To be able to compare the parameters $S_0, \ Q \ \text{and} \ \omega_0$ present in the kernels with the characteristic quantities found for stellar signals in power-spectrum fitting, a correspondence needs to be established between the functions used to capture the signals in both domains.

Concerning granulation, Eq.~(\ref{eq_psd_gran}) can be related to the one presented in \citet{kallinger_2014}:
\begin{equation}
    S(\nu) = \frac{2\sqrt{2}}{\pi} \frac{a_\text{gran}^2/b_\text{gran}}{\left( \nu/b_\text{gran} \right)^4 +1} \, ,
\label{eq_psd_kallinger_gran}
\end{equation}
where $a_\text{gran}$ and $b_\text{gran}$ are the characteristic amplitude and frequency of the granulation signal, respectively. Note that the above PSD varies in linear frequency, $\nu$. For the oscillation bump, which is commonly modelled as a Gaussian in a power spectrum analysis, Eq.~(\ref{eq_psd_sho}) can be related to \citet{kallinger_2014}
\begin{equation}
    S(\nu) = P_g \ \text{exp} \left( \frac{-(\nu - \nu_\text{max})^2}{2 \sigma^2} \right) \, ,
    \label{eq_psd_kallinger_bump}
\end{equation}
where $P_g$ is the height of the oscillation bump, $\nu_\text{max}$ is the frequency of maximum oscillation amplitude, and $\sigma$ is the width of the bump. Again, note that this PSD varies in linear frequency.

An illustrative example of the power spectra of both kernels described above is shown in Fig.~\ref{fig:comparison_kernel_psd}. It shows the shape of the granulation PSD (Eq.~\ref{eq_psd_gran}), common to both the GP and power spectrum analyses, while highlighting the differences in the functions used to capture the signal from the oscillation bump, namely, the low-frequency tail and the slow decrease in power at high frequencies in the GP model.

In order to make a meaningful comparison between the parameters in Eqs.~(\ref{eq_psd_gran}) and (\ref{eq_psd_sho}) and those in Eqs.~(\ref{eq_psd_kallinger_gran}) and (\ref{eq_psd_kallinger_bump}), respectively, all equations need to be consistently normalized. Appendix \ref{append:normal} details the steps taken to determine the exact correspondence between the parameters in both approaches. After applying a constant normalization factor to both Eqs.~(\ref{eq_psd_kallinger_gran}) and (\ref{eq_psd_kallinger_bump}), the parameters returned by the two methods can be compared:
\begin{align}
    \begin{split}
        a_\text{gran}  \ = \ & \sqrt{\sqrt{2} \ S_{0,\text{gran}} \ \omega_{0,\text{gran}}} \; ,	\\
        b_\text{gran}  \ = \ & \frac{\omega_{0,\text{gran}}}{2\pi} \; , 				\\
        P_g            \ = \ & 4 \ S_{0,\text{bump}} \ Q_\text{bump}^2 \; ,			\\
        \nu_\text{max} \ = \ & \frac{\omega_{0,\text{bump}}}{2\pi} \; . 				\\
    \end{split}
\end{align}
The subscripts ``gran'' and ``bump'' emphasize that, while the parameter names $S_0, \ Q \ \text{and} \ \omega_0$ are the same for both kernels, these are separate kernels in the model, and thus have different values.

\subsection{Parameter estimation}
\label{sec:priors}

To both determine the most appropriate value for each parameter in a model and also study its uncertainties, the {\sc ecmee}\footnote{\url{https://github.com/dfm/emcee}} {\sc Python} package is used. This package provides Bayesian parameter estimation given a prior for each of the parameters in the model.

In this work, only uniform priors have been considered. Tables \ref{priors_model1} and \ref{priors_model2} present the lower and upper bounds chosen for the prior distributions of all parameters in Models 1 and 2, respectively.

\begin{table}
	\centering
	\begin{tabular}{l r r}
		\hline
		Parameter                   & Lower Bound & Upper Bound \\
		\hline
		$a_\text{gran,1}$ (ppm)     & 10          & 400         \\
		$b_\text{gran,1}$ ($\mu$Hz) & 10          & 200         \\
		$P_\text{g}$ (ppm)          & 10          & 1800        \\
		$Q_\text{bump}$             & 1.2         & 18          \\
		$\nu_\text{max}$ ($\mu$Hz)  & 80          & 220         \\
		White Noise (ppm)           & 0           & 400         \\
		\hline
	\end{tabular}
	\caption{Lower and upper bounds chosen for the uniform distributions used as priors for parameters in Model 1.}
	\label{priors_model1}
\end{table}

\begin{table}
	\centering
	\begin{tabular}{l r r}
		\hline
		Parameter                   & Lower Bound & Upper Bound \\
		\hline
		$a_\text{gran,1}$ (ppm)     & 10          & 400         \\
		$b_\text{gran,1}$ ($\mu$Hz) & 10          & 70          \\
		$a_\text{gran,2}$ (ppm)     & 10          & 400         \\
		$b_\text{gran,2}$ ($\mu$Hz) & 80          & 300         \\
		$P_\text{g}$ (ppm)          & 10          & 1800        \\
		$Q_\text{bump}$             & 1.2         & 18          \\
		$\nu_\text{max}$ ($\mu$Hz)  & 80          & 220         \\
		White Noise (ppm)           & 0           & 400         \\
		\hline
	\end{tabular}
	\caption{Lower and upper bounds chosen for the uniform distributions used as priors for parameters in Model 2.}
	\label{priors_model2}
\end{table}

The parameter space was explored by $w$ Monte Carlo Markov Chains where each chain produced 25000 samples. The number of chains $w$ was defined as 4 times the number of free parameters in the chosen model.

To ensure convergence of the chains, a range of tests were performed on the samples obtained. The Geweke statistic \citep{geweke_1992} was applied to all the chains to estimate the burn-in period of each chain. Chains that did not pass this test were removed from the determination of the final parameters. In the next step, chains with low posterior probabilities were discarded (average posterior probability of samples in chain lower than the 10th percentile of the posterior probability distribution of all samples) in order to remove any chains converging to local minima.

From the remaining chains, the univariate and multivariate Gelman$-$Rubin diagnostics \citep{gelman_1998} were determined, which evaluate convergence of Markov chains by comparing the between-chains and within-chain variance. Whilst the univariate approach analyses each parameter of the model independently, the multivariate version of this diagnostic takes into account covariances between the parameters, so it is a more demanding diagnostic for convergence. While these tests are not strictly valid to apply to these chains, since {\sc ecmee} produces correlated chains, the diagnostics still provide a rough numerical evaluation of convergence of the results. 
When determining both of these diagnostics, up to 50\% of the samples from each chain were allowed to be discarded in order to improve the convergence of the remaining samples. An initial maximum value of 1.1 was defined as the threshold enforced for the Gelman-Rubin diagnostic.

The tests applied showed that, after selection of the best samples and chains, even in the rare event where only 25\% of the initial samples where considered, the number of samples was large enough to achieve accurate estimations on all parameters.


Taking the final selected samples, both the median and mode were calculated for each of the model's parameters along with two highest posterior density (HPD) intervals. The first HPD interval calculated was a 68.3\% HPD which was used as a measure of the lower and upper uncertainty on the value of each parameter (so that meaningful uncertainties could also be determined for non-normal sample distributions). The second was a 95\% HPD interval which was compared with the width of the uniform prior defined for each of the model's parameters in order to determine whether the converged samples were exploring a small subset of the prior space or if the entire prior had similar posterior probability, in which case the specific parameter was flagged as not capable of being constrained.

\section{Application to TESS-like artificial data}\label{sec:appTESS}
\subsection{Methodology}\label{sec:tess_methodology}

The method is first tested with TESS-like artificial time series for a set of 20 LLRGB stars (with effective temperature $4800 < T_\text{eff} < 5500\:\text{K}$, frequency of maximum oscillation amplitude $105 < \nu_\text{max} < 185\:\mu\text{Hz}$, and apparent magnitude $V < 11$). Generation of the artificial light curves is performed originally in the frequency domain by using scaling relations, after which an inverse Fourier transform is applied and the 30-min cadence of TESS full-frame images considered. A photometric noise model for TESS \citep{sullivan_2015,campante_2016b} is used in order to predict the rms noise for a given exposure time. A systematic noise term of $20\:\text{ppm}\,\text{hr}^\text{1/2}$ was included in this calculation. To model the granulation power spectral density, a scaled version (to predict TESS granulation amplitudes) of model F of \citet{kallinger_2014} was adopted, which contains two granulation (or Harvey-like) components, the granulation and mesogranulation. No aliased granulation power was considered. Individual radial, (mixed) dipole and quadrupole modes were also modelled \citep{kuszlewicz_2019}. 

For each of the 20 simulated stars, 10 independent 27.4-day time series were generated. The values and corresponding uncertainties computed for each of the model parameters take into account the analysis of at least 5 and up to 10 of these independent time series (chosen according to their performance in the evaluated convergence statistics, as detailed in Section~\ref{sec:priors}) so as to reduce the possibility of systematic errors. Specifically, each parameter was estimated as the median of the values obtained for the chosen independent time series. Its uncertainty was defined as the sum in quadrature of the uncertainty associated with the median value and the standard deviation of the values determined for the remaining chosen time series. Furthermore, if any of the parameters from the chosen runs are flagged as not constrained, the final parameter calculated from these runs will also be flagged.

\subsection{Results}\label{sec:tess_results}

\begin{figure*}
	\centering
	\includegraphics[width=0.79\textwidth]{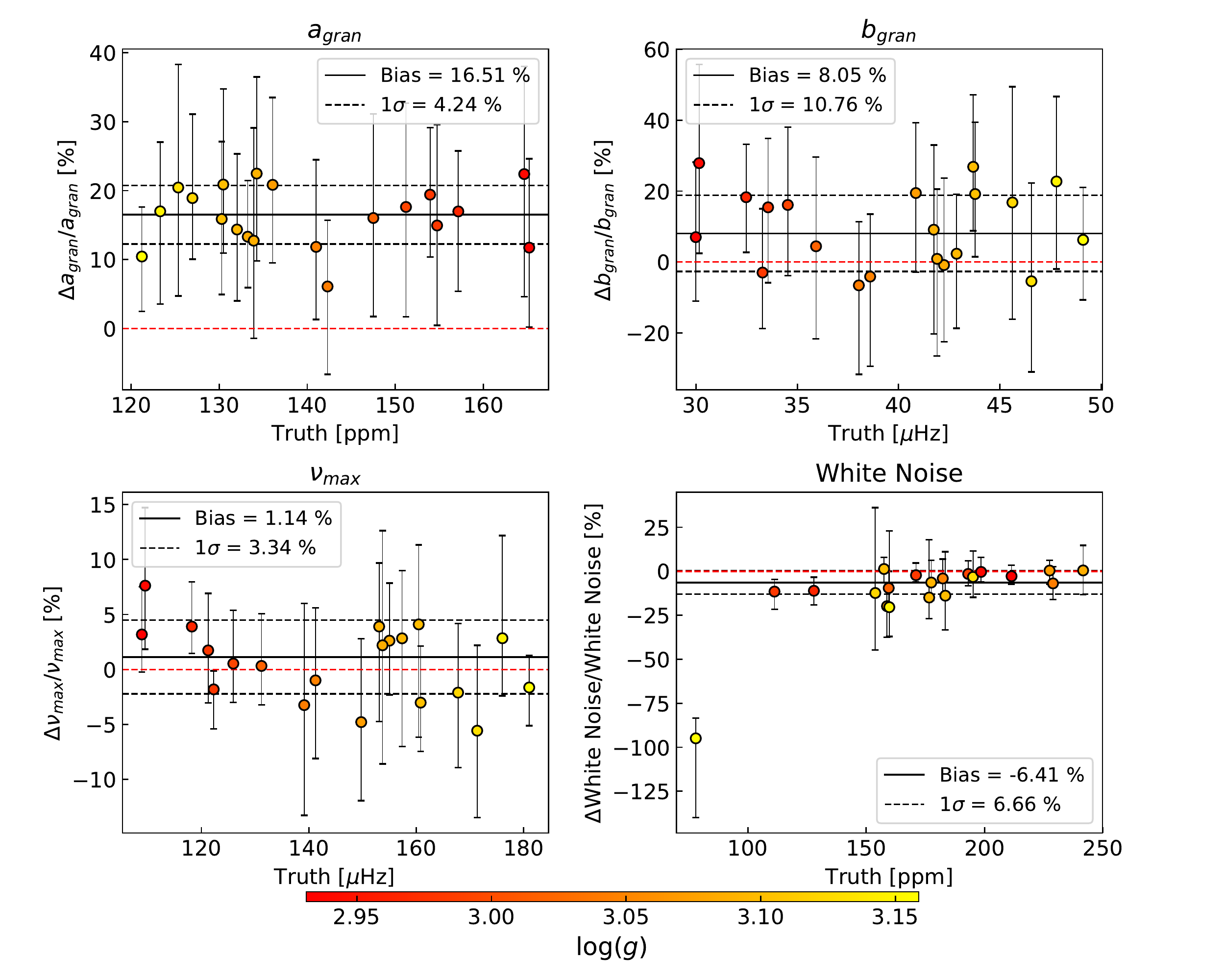}
	\caption{Comparison of the parameters in the fit of Model 1 to the TESS-like artificial data with the input used to generate those data. Data points represent the relative deviation with respect to the input value, with error bars corresponding to the uncertainties returned by the GP regression method. Black solid and dashed lines represent the median and standard deviation of the data points, respectively, with their numerical values shown in the inset. The red dashed line denotes a null offset. Data points are colour-coded according to a star's surface gravity, $\log g$.}
	\label{fig:comparison_tess_model1}
\end{figure*}

\begin{figure*}
	\centering
	\includegraphics[width=0.79\textwidth]{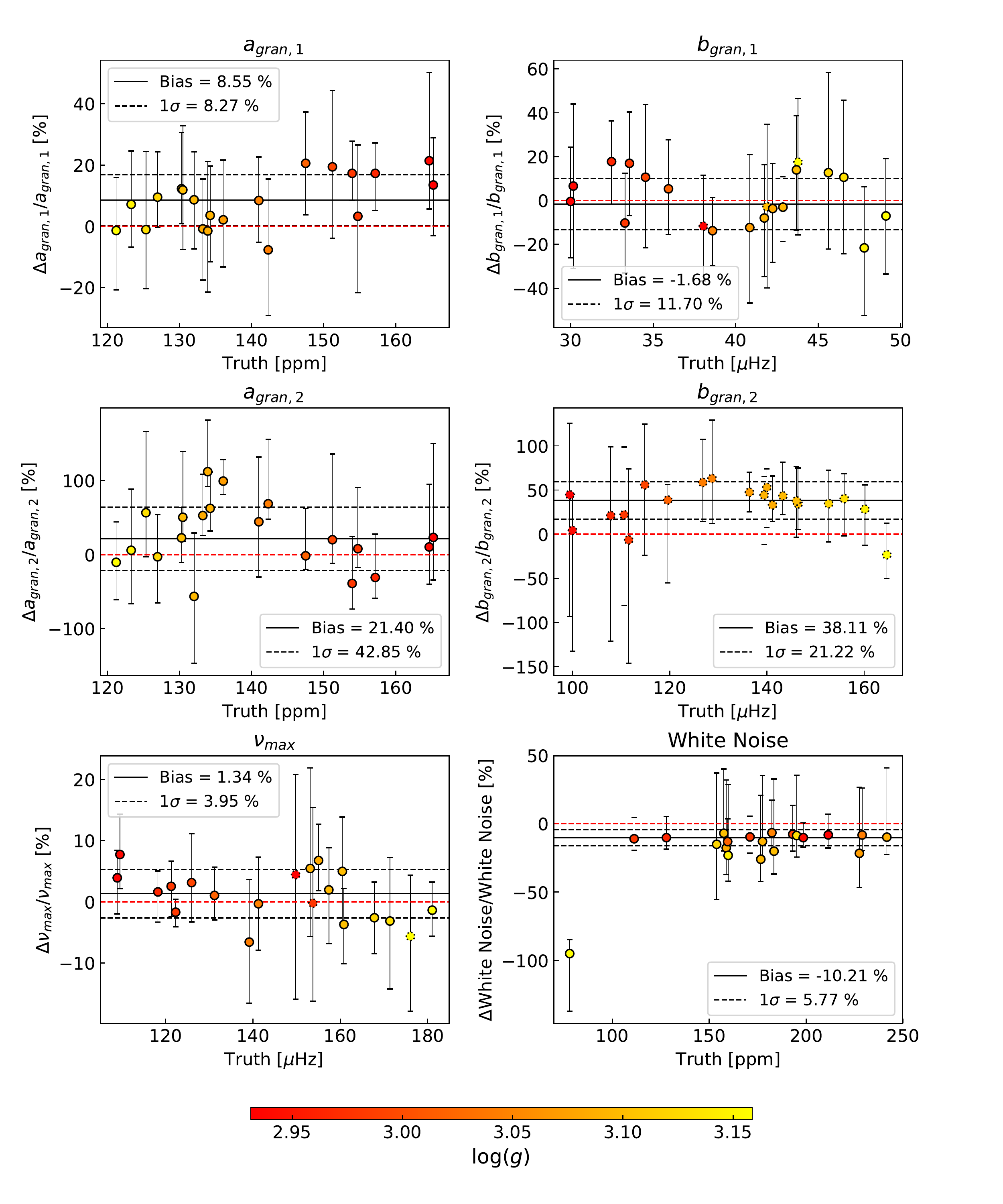}
	\caption{Comparison of the parameters in the fit of Model 2 to the TESS-like artificial data with the input used to generate those data. Data points represent the relative deviation with respect to the input value, with error bars corresponding to the uncertainties returned by the GP regression method. Parameters that have been flagged as not constrained (see end of Section~\ref{sec:priors}) have dotted edges. Black solid and dashed lines represent the median and standard deviation of the data points, respectively, with their numerical values shown in the inset. The red dashed line denotes a null offset. Data points are colour-coded according to a star's surface gravity, $\log g$.}
	\label{fig:comparison_tess_model2}
\end{figure*}

\begin{figure}
	\centering
	\includegraphics[width=\columnwidth]{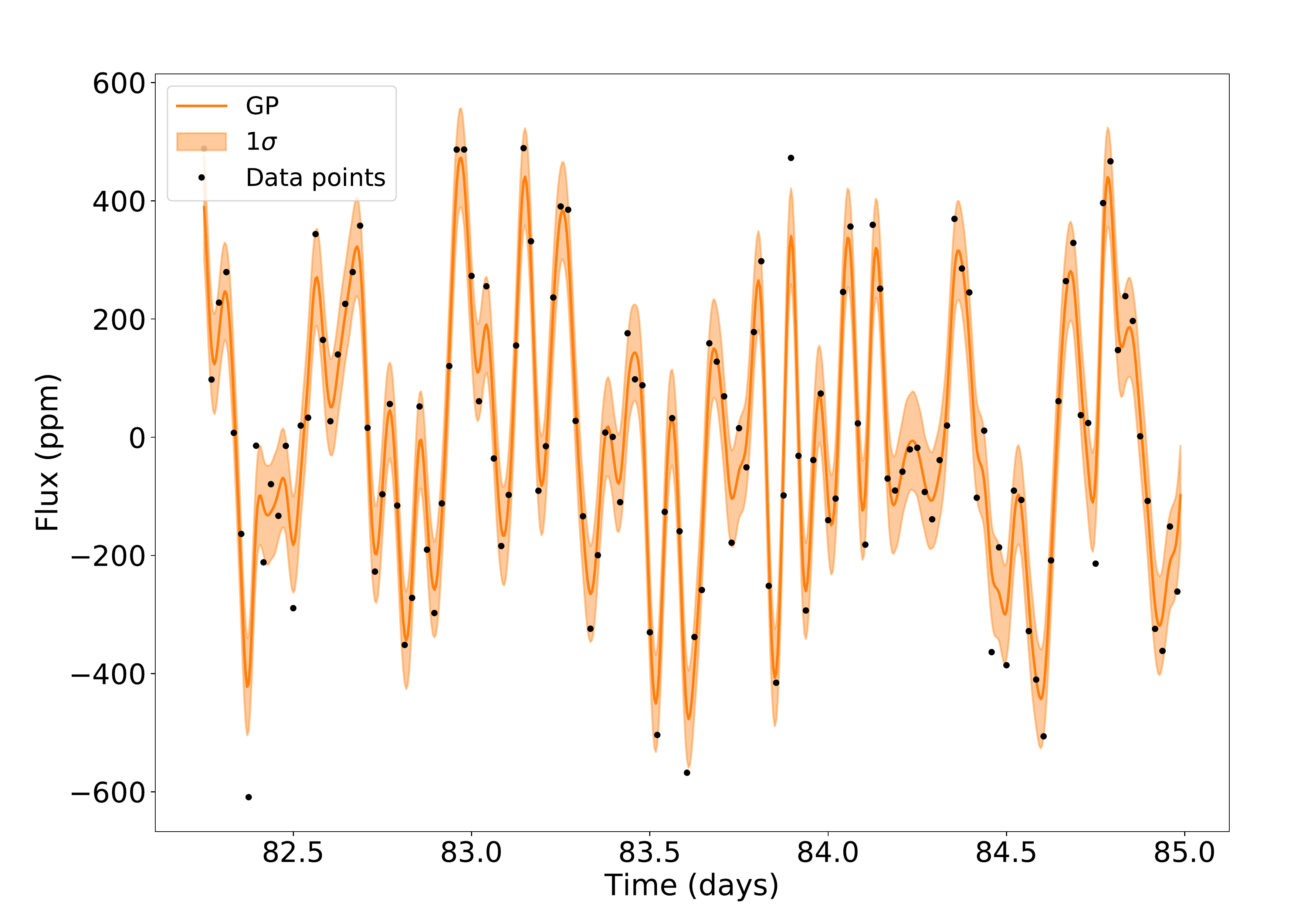}
	\caption{Predictive model (Model 1) output by the GP regression (mean and $1 \sigma$ interval) when applied to one of the artificial TESS-like time series. The plot is zoomed in on the first $\sim 3$ days of simulated data to improve visualization.}
	\label{fig:tess_gp_zoom}
\end{figure}

\begin{figure}
	\centering
	\includegraphics[width=\columnwidth]{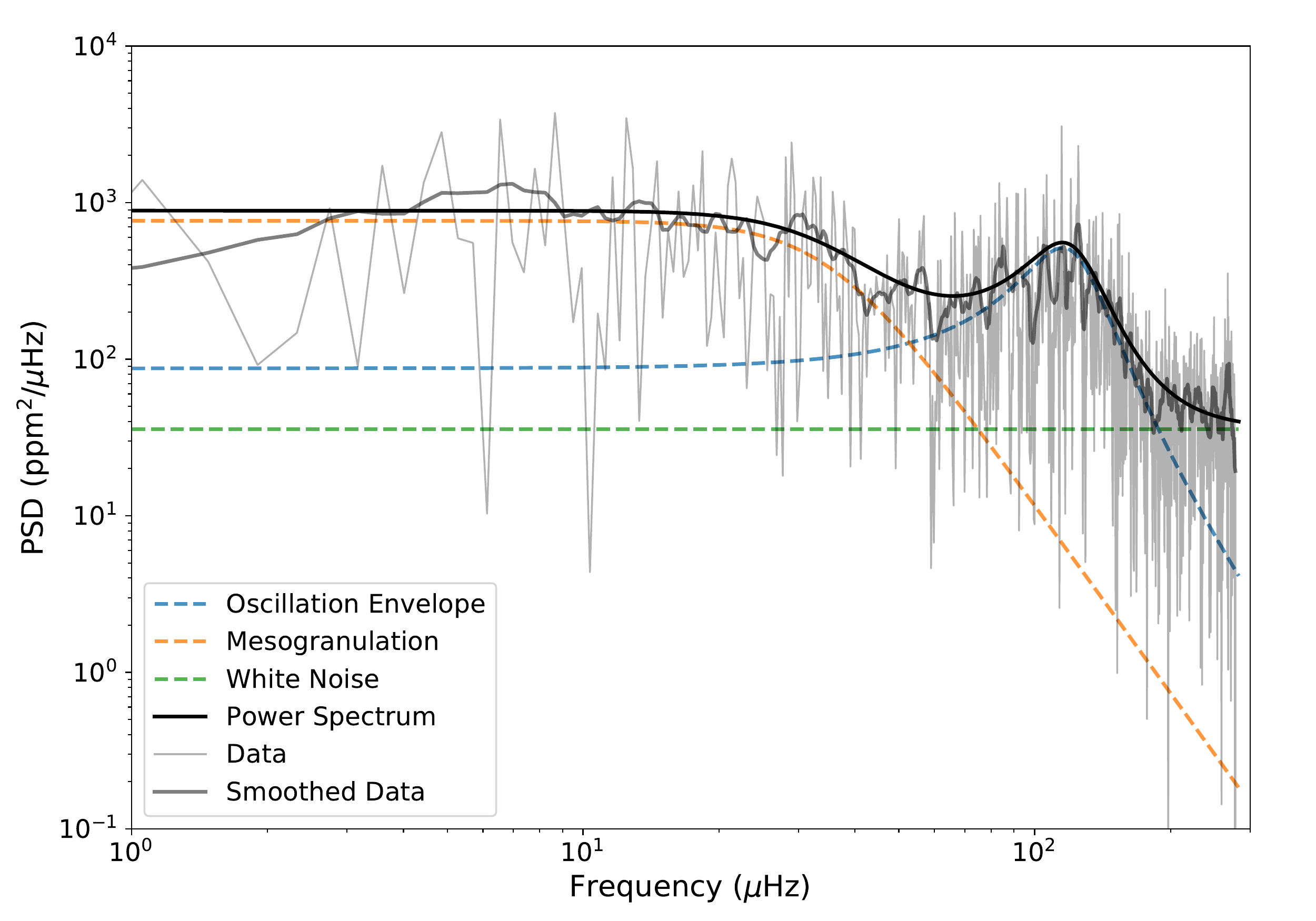}
	\caption{Power spectral density of the same (full) light curve depicted in Fig.~\ref{fig:tess_gp_zoom}. The PSD of the light curve is shown in light grey, with a slightly smoothed version overlapped in dark grey. The PSD of the GP regression output (Model 1) is shown as a solid red curve, with individual components shown in different line styles and colors (see legend).}
	\label{fig:tess_psd}
\end{figure}

The method is applied twice to all data sets, making use of each of the models, i.e., Models 1 and 2. The parameters derived from the GP regression may then be compared to the input parameters used when generating the light curves. Figures \ref{fig:comparison_tess_model1} and \ref{fig:comparison_tess_model2} show such a comparison for the parameters of Models 1 and 2, respectively. The parameters represented in Fig.~\ref{fig:comparison_tess_model1} for Model 1 are the amplitude  and characteristic frequency of the mesogranulation component in the model, $a_\text{gran,1}$ and $b_\text{gran,1}$, respectively, the frequency of maximum oscillation, $\nu_\text{max}$, and the white noise. Figure \ref{fig:comparison_tess_model2} depicts the same parameters with the addition of the amplitude and characteristic frequency of the granulation component, $a_\text{gran,2}$ and $b_\text{gran,2}$, respectively. Both the oscillator's quality factor, $Q$, and the power at $\nu\!=\!\nu_\text{max}$, $P_\text{g}$, are not represented as the artificial light curves result from simulating individual oscillation modes and hence there are no {\it bona fide} input values to compare with.

As an example of how well GP regression allows for the characterization of the stellar signal in the time domain, Fig.~\ref{fig:tess_gp_zoom} shows a blowup of the fit performed to one of the artificial time series. Figure \ref{fig:tess_psd} shows the PSD of that same GP regression output compared to the PSD of the light curve. Both figures show the results obtained when fitting Model 1 to the data.

In what follows, a given parameter is considered to have been accurately determined if the null offset (red dashed line) is within the $1\sigma$ interval (black dashed lines) associated with the median of the data points (or bias; black solid line). Looking at the results obtained when employing Model 1 (Fig.~\ref{fig:comparison_tess_model1}), the amplitude, $a_\text{gran,1}$, of the mesogranulation component in the model is not correctly retrieved, showing a bias of 16.51\% (4.24\% scatter) relative to the input values in the mesogranulation component in the data, whilst the characteristic frequency, $b_\text{gran,1}$, is correctly recovered with a bias of 8.05\% (10.76\% scatter) This is not too surprising, since the model being considered is incomplete: the mesogranulation amplitude is being overestimated in an attempt to capture the power in the two granulation components present in the data. It should be noted that the low-frequency tail of the oscillation bump profile (see Sect.~\ref{sec:implement} and Fig.~\ref{fig:comparison_kernel_psd}) contributes to somewhat attenuating this offset. Nevertheless, the estimation of $\nu_\text{max}$ is robust, with this parameter being accurately (1.14\% bias) and precisely (3.34\% scatter) recovered. Finally, the white noise level is recovered to within 7\% of the input value. The slight, overall underestimation of the white noise level is to be expected because of the non-negligible contribution of the oscillation bump profile at high frequencies (see Fig.~\ref{fig:comparison_kernel_psd}). Concerning the outlying artificial star, it has the highest value of $\nu_\text{max}$ amongst the stars in the sample. Upon inspection of its PSD, it becomes clear that the proximity of the oscillations to the Nyquist frequency ($\nu_\text{Nyq}\!\approx\!283\: \mu$Hz for the 30-min cadence of the simulated light curves) prevents the white noise level from being robustly determined. For this reason, this star was not considered when determining the bias and scatter for the white noise comparison.

Concerning Model 2 (Fig.~\ref{fig:comparison_tess_model2}), the introduction of a second granulation component in the model leads to an improvement in the fit to the mesogranulation signal, with the amplitude and characteristic frequency within 8.55\% (8.27\% scatter) and $-$1.68\% (11.70\% scatter) of the input values, respectively. Results are, however, noticeably less robust for the added granulation component, with the correct amplitude of the granulation being within uncertainties only due to the high scatter (21.40\% bias and 42.85\% scatter) and the characteristic frequency not being constrained at all (38.11\% bias and 21.22\% scatter). Finally, the estimation of $\nu_\text{max}$ continues to be robust (1.34\% bias and 3.95\% scatter) whilst the white noise level could not recovered ($-$10.61\% bias and 5.77\% scatter).

All in all, the introduction of the second granulation component does improve the fit to the mesogranulation but the granulation's characteristic frequency cannot be constrained for any star and both the characteristic frequency of the mesogranulation and $\nu_\text{max}$ show unconstrained results for some of the stars. The white noise is also not recovered when adding the extra component to the model. Considering the high white noise levels expected for TESS data, as well as the short duration of typical TESS time series, Model 1 with only the mesogranulation seems to be better suited to accurately find the stellar signals of RGB stars observed by TESS.


\section{Application to \textit{Kepler} LLRGB stars}\label{sec:appKepler}
\subsection{Methodology}\label{sec:kepler_methodology}

As a second test, the method is applied to the same sample of 19 \textit{Kepler} LLRGB stars considered in \citet{corsaro_2015b}. In order to mimic the typical amount of data expected from TESS and to account for systematics that might be present in the light curve, 10 non-overlapping subsets of 27.4 days of observations were considered from the full \textit{Kepler} light curve for each star.

The values and corresponding uncertainties computed for each of the model parameters again take into account the analysis of at least 5 and up to 10 of these independent subsets, as described in the previous section. 


Besides performing a GP regression in the time domain, a standard fit to the power spectrum was also conducted making use of a model similar to the GP kernel (see Section \ref{sec:implement}). The latter analysis was performed using the {\sc diamonds} code\footnote{\url{https://github.com/EnricoCorsaro/DIAMONDS}} \citep{corsaro_2014}, which fits the power spectrum and determines the model parameters within a Bayesian framework. 


\subsection{Results}\label{sec:kepler_results}

\begin{figure*}
	\centering
	\includegraphics[width=0.79\textwidth]{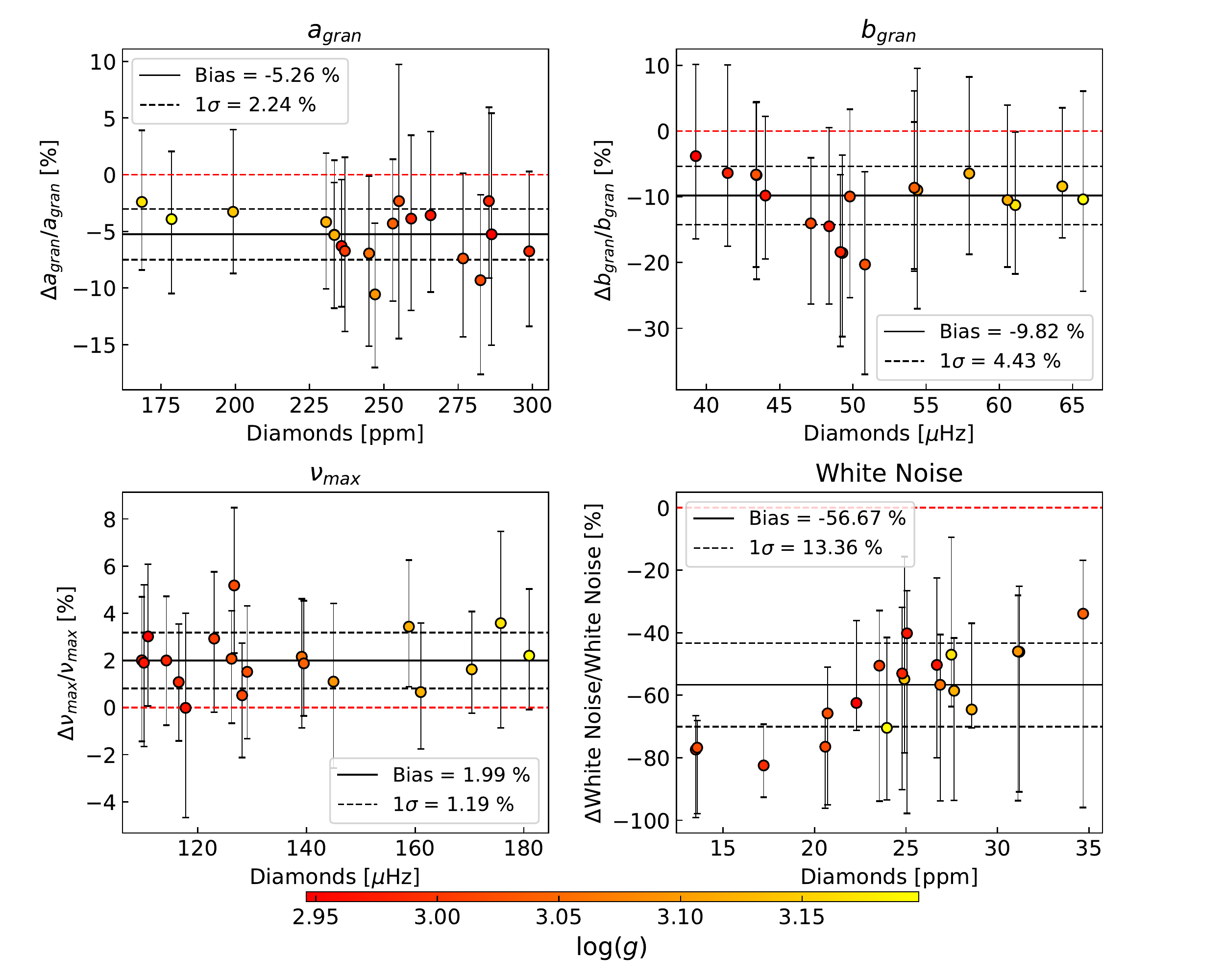}
	\caption{Comparison of the parameters in the fit of Model 1 to the \textit{Kepler} time series data both by means of a GP regression and power-spectrum fitting. Data points represent the relative deviation with respect to the value determined using the PSD-fitting procedure, with error bars corresponding to the sum in quadrature of the uncertainties of both methods. Black solid and dashed lines represent the median and standard deviation of the data points, respectively, with their numerical values shown in the inset. The red dashed line denotes a null offset. Data points are colour-coded according to a star's surface gravity, $\log g$.}
	\label{fig:comparison_diamonds_model1}
\end{figure*}

\begin{figure*}
	\centering
	\includegraphics[width=0.79\textwidth]{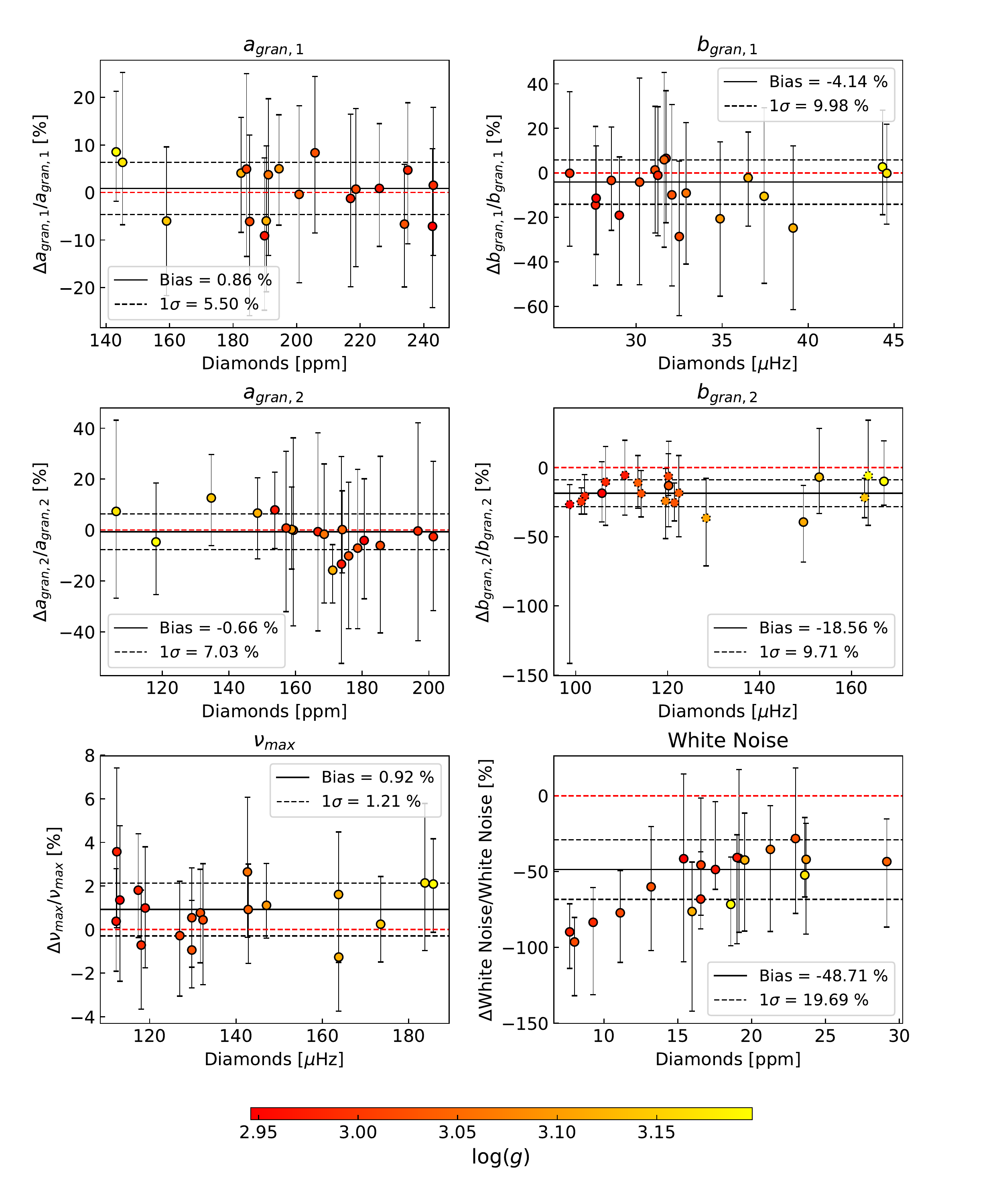}
	\caption{Comparison of the parameters in the fit of Model 2 to the \textit{Kepler} time series data both by means of a GP regression and power-spectrum fitting. Data points represent the relative deviation with respect to the value determined using the PSD-fitting procedure, with error bars corresponding to the sum in quadrature of the uncertainties of both methods. Parameters that have been flagged as not constrained (see end of Section~\ref{sec:priors}) have dotted edges. Black solid and dashed lines represent the median and standard deviation of the data points, respectively, with their numerical values shown in the inset. The red dashed line denotes a null offset. Data points are colour-coded according to a star's surface gravity, $\log g$.}
	\label{fig:comparison_diamonds_model2}
\end{figure*}

\begin{figure}
	\centering
	\includegraphics[width=\columnwidth]{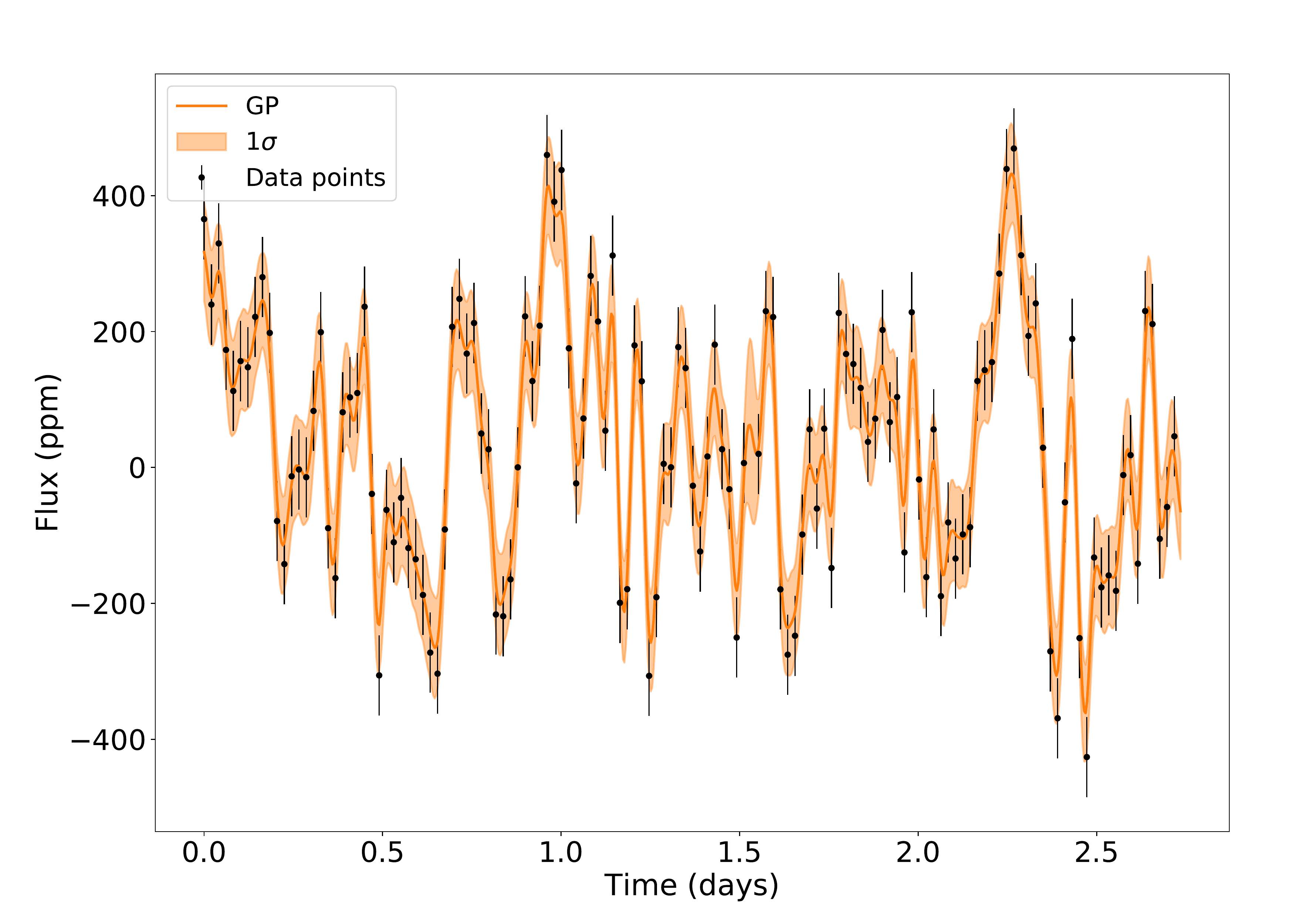}
	\caption{Predictive model (Model 2) output by the GP regression (mean and $1\sigma$ interval) when applied to one of the \textit{Kepler} LLRGB stars in the sample. The plot is zoomed in on the first $\sim3$ days of observations to improve visualization.}
	\label{fig:diamonds_gp_zoom}
\end{figure}

\begin{figure}
	\centering
	\includegraphics[width=\columnwidth]{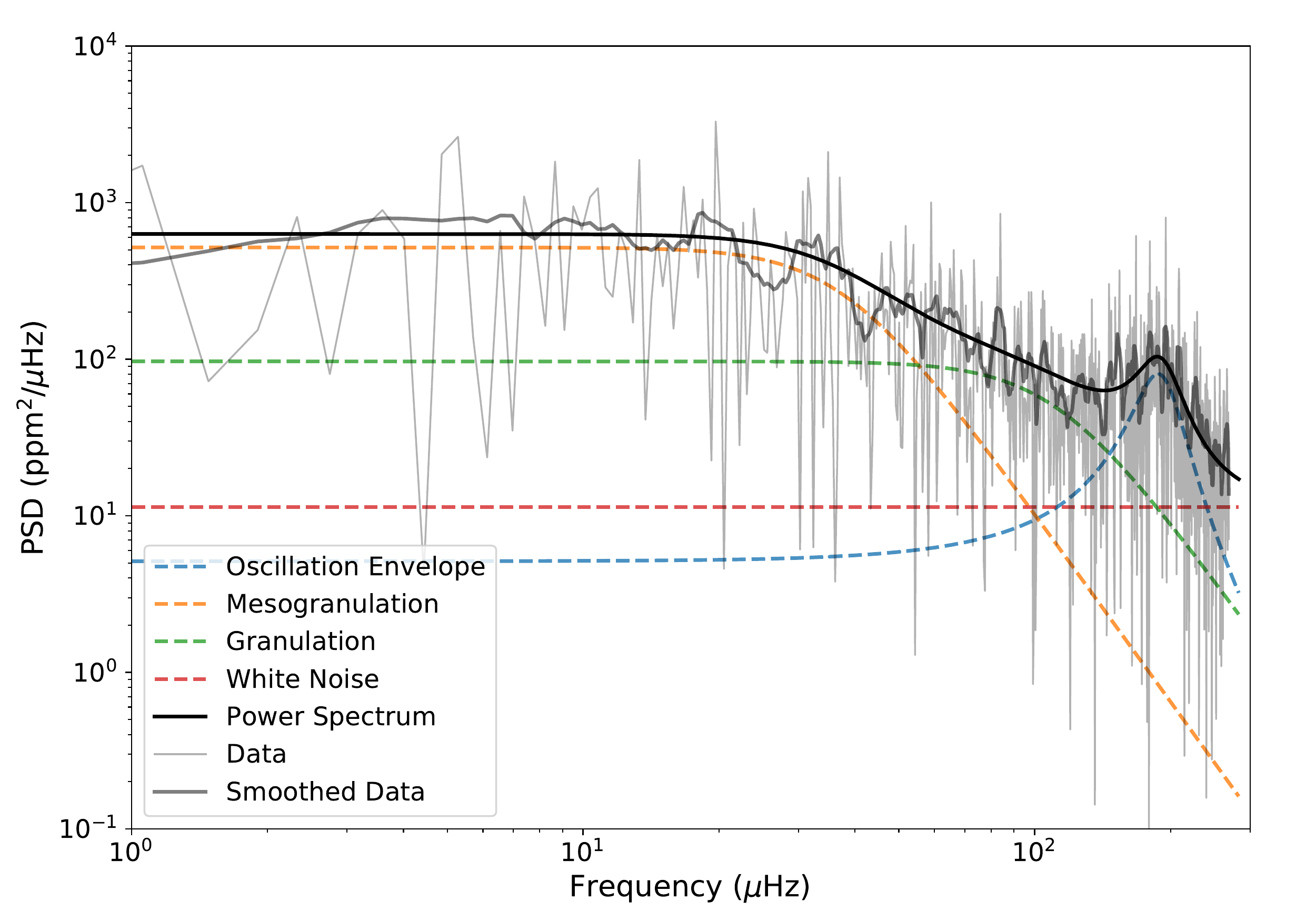}
	\caption{Power spectral density of the same (full) light curve depicted in Fig.~\ref{fig:diamonds_gp_zoom}. The PSD of the light curve is shown in light grey, with a slightly smoothed version overlapped in dark grey. The PSD of the GP regression output (Model 2) is shown as a solid red curve, with individual components shown in different line styles and colors (see legend).}
	\label{fig:diamonds_psd}
\end{figure}

A comparison of the output obtained with the time domain GP and the power spectrum fit is shown in Figs.~\ref{fig:comparison_diamonds_model1} and \ref{fig:comparison_diamonds_model2} for Models 1 and 2, respectively. The parameters depicted are the same as in Figs.~\ref{fig:comparison_tess_model1} and \ref{fig:comparison_tess_model2}. A blowup of the output of the GP regression when applied to one of the \textit{Kepler} LLRGB stars in the sample is shown in Fig.~\ref{fig:diamonds_gp_zoom}. Figure \ref{fig:diamonds_psd} shows the PSD of that same GP regression output compared to the PSD of the light curve. Both figures show the results obtained when fitting Model 2 to the data.

Contrary to the previous section, where the method's accuracy was assessed with artificial time series, the method is now tested on real \textit{Kepler} data for a sample of well-studied LLRGB stars. By doing this, the parameters derived through GP regression can be compared to the equivalent parameters obtained when performing a standard fit to the power spectrum. This test thus allows for a comparison with the more traditional methodology used in studies of stellar light curves.

Looking at the results obtained when considering Model 1 (Fig.~\ref{fig:comparison_diamonds_model1}), the parameters describing the mesogranulation component in the model, $a_\text{gran,1}$ and $b_\text{gran,1}$, are underestimated, with relative biases of about $-$5\% and $-$10\%, respectively. This underestimation is expected due to the presence of the low-frequency tail of the oscillation bump profile (see Fig.~\ref{fig:comparison_kernel_psd}). To confirm that these offsets between parameters were only due to the differences in the models considered (specifically the model that captures the signal from the oscillations), a second fit to the power spectrum was performed, where the equations chosen were the exact power spectrum equations of the kernels used in the GP model. This test confirmed that, when the models are an exact match, both the GP regression and the frequency domain fit recover the same results. 

Concerning $\nu_\text{max}$, a small relative bias of 1.99\% (1.19\% scatter) is found between methods. Finally, results for the white noise level show a relative bias of about $-$56\% (13.36\% scatter) between the two methods. It should be borne in mind that white noise levels for the \textit{Kepler} stars in the sample are relatively low (\textit{Kepler}'s effective collecting area is larger than that of the individual TESS cameras by a factor of $\sim\!10^2$), which, coupled with differences in the models adopted in either method (see Fig.~\ref{fig:comparison_kernel_psd}), results in large relative differences. Inspection of the absolute value of this same bias reveals differences no greater than 15 ppm and similar between stars, which results in the trend seen in the relative offsets, where stars with higher white noise (as determined by Diamonds) have a lower relative offset.

With respect to Model 2 (Fig.~\ref{fig:comparison_diamonds_model2}), excellent agreement is seen between the parameters describing the mesogranulation component, i.e., $a_\text{gran,1}$ and $b_\text{gran,1}$, with relative differences of 0.86\% and $-$4.14\%, respectively. The low-frequency tail of the oscillation bump profile does not appear to be affecting this component. It does, however, impact the parameters describing the second granulation component. Whilst the amplitude of the component, $a_\text{gran,2}$, is accurately recovered with a bias of $-$0.66\% (7.03\% scatter), the characteristic frequency is systematically shifted, showing a bias of $-$18.56\% (9.71\% scatter), with more than half the stars having parameters that have not been well constrained (dotted edges). Regarding $\nu_\text{max}$, just like with Model 1, a small relative bias of 0.92\% (1.21\% scatter) is found between methods. For the white noise level, results show a less pronounced discrepancy compared to the one seen for Model 1, with the relative bias between methods now being of $-$48.71\% (16.89\% scatter) and the absolute difference never exceeding 6 ppm.


\subsection{Uncertainties in $\nu_{\text{max}}$}\label{sec:nu_max_uncertainties}

Taking the uncertainties determined for the estimation of $\nu_\text{max}$ by both the GP method and the {\sc diamonds} code, these can be compared to an expected value for these uncertainties. Taking into account results from asteroseismology of red giants from the first four months of \textit{Kepler} data, \citet{kallinger_2010} defined the following relation to determine $\sigma_{\nu_\text{max}}$, a lower limit for the uncertainty in $\nu_\text{max}$:
\begin{equation}
	\label{eq_sigma_numax}
	\sigma_{\nu_\text{max}} = \nu_\text{res} \left( 1 + \frac{4}{(HBR/\sigma_\text{g})^{2/3}} \right),
\end{equation}
where \textit{HBR} is the height-to-background ratio, defined as the ratio between the power of the oscillation bump, $P_\text{g}$, and the background signal at $\nu\!=\!\nu_\text{max}$, $B_{\nu_\text{max}}$, $\sigma_\text{g}$ is the width of the oscillation bump, and $\nu_\text{res}$ is the frequency resolution, which is the inverse of the data set length.

Estimates of $P_\text{g}$, $B_{\nu_\text{max}}$ and $\sigma_\text{g}$ can be obtained from the scaling relations found in \citet{mosser_2012a},
\begin{alignat}{2}
	& P_\text{g} = 2.03 \times 10^7 \times \nu_\text{max}^{-2.38}, \\
	& B_{\nu_\text{max}} = 6.37 \times 10^6 \times \nu_\text{max}^{-2.41},
\end{alignat}
and in \citet{campante_2016b},
\begin{alignat}{2}
	& \text{FWHM}_\text{g} = \frac{\nu_\text{max}}{2}, \\
	& \sigma_\text{g} = \frac{\text{FWHM}_\text{g}}{2 \sqrt{2 \ \text{ln}(2)}},
\end{alignat}
where $\text{FWHM}_\text{g}$ is the full width at half maximum of the oscillation bump.

Using the values of $\nu_\text{max}$ obtained by the GP method, Eq.~(\ref{eq_sigma_numax}) can be used to calculate an estimate of $\sigma_{\nu_\text{max}}$, which can be compared to the uncertainties in the determination of $\nu_\text{max}$ (considering Model 1) calculated by both the GP method, $\sigma_\text{GP}$, and the {\sc diamonds} code, $\sigma_\text{Diamonds}$. The average absolute and relative values obtained for these quantities considering all 19 stars in the sample are:
\begin{itemize}
	\item $\left \langle \sigma_{\nu_\text{max}} \right \rangle \!=\! 7.15 \ \mu$Hz, $(5.20\%)$
	\item $\left \langle \sigma_\text{GP} \right \rangle \!=\! 3.43 \ \mu$Hz, $(2.46\%)$
	\item $\left \langle \sigma_\text{Diamonds} \right \rangle \!=\! 1.91 \ \mu$Hz, $(1.44\%)$
\end{itemize}
The values above would suggest that both methods applied here are underestimating the uncertainties in $\nu_\text{max}$. 
However, {\sc diamonds} has successful applications in the literature, including with TESS observations \citep{corsaro_2015b,huber_2019}, which suggests that the determination of its uncertainties is correct (see Sec.~4.5 in \citet{corsaro_2014} for uncertainty estimation). 
The uncertainty determination in the GP method was described in detail in Sec.~\ref{sec:priors}, where the MCMC sampling was done with {\sc emcee}, and the obtained values are similar to those of {\sc diamonds}.
Since the model from Eq.~(\ref{eq_sigma_numax}) was determined based on the uncertainties estimated by \citet{kallinger_2010}, following the methodology described in \citet{gruberbauer_2009}, the difference between the expected uncertainty, $\sigma_{\nu_\text{max}}$, and the calculated uncertainties, $\sigma_\text{GP}$ and $\sigma_\text{Diamonds}$, might be related to the different methods adopted.

\section{Exoplanet transits}\label{sec:transit}

To test the applicability of the method in the context of exoplanet transit modelling, a transit model \citep{kreidberg_2015} was added to the GP model presented in this article. This new combined model should be capable of capturing both the stellar and the transit signals simultaneously when modelling the light curve of a star.

Simulated transits of giant planets were injected into the sample of simulated light curves from Sec.~\ref{sec:appTESS} (see Sec.~3.1 from \citet{campante_2018}). The detection of the injected transits was then evaluated using the BLS (Box-fitting Least Squares) method \citep{kovacs_2002}, and a signal detection efficiency (SDE) threshold was defined, being that all detections with an SDE above this threshold were considered likely planetary transit detections. Finally, for all likely detections, the light curves were modelled using the new combined model (with the stellar GP model considered following Model 1 introduced in Sec.~\ref{sec:implement}) as well as a simple transit model with white noise. This second, simpler model was considered in order to evaluate the improvement in the determined transit parameters when modelling the stellar signals simultaneously with the transit.

\begin{figure}
	\centering
	\includegraphics[width=\columnwidth]{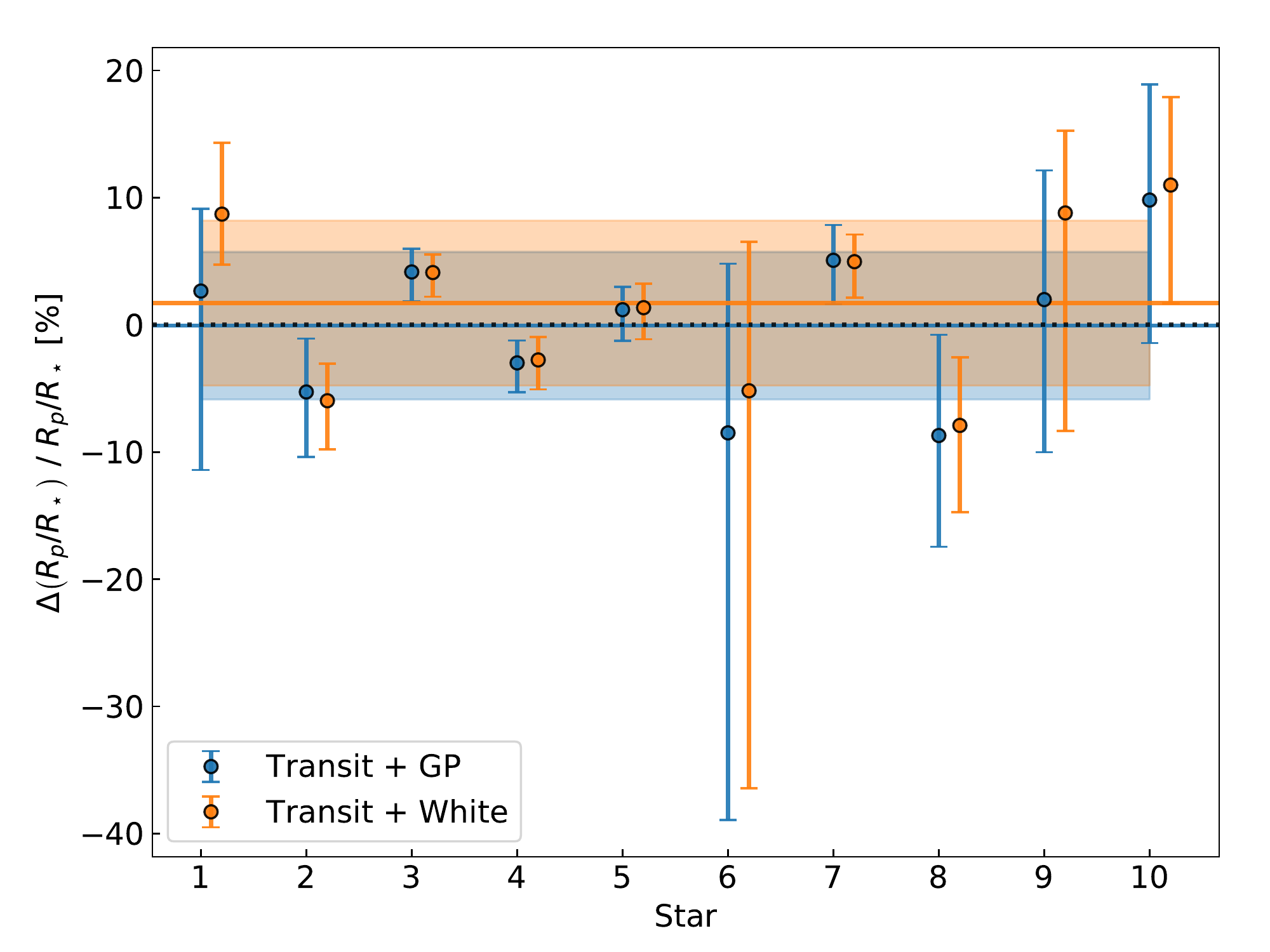}
	\caption{Comparison of the ratio between the planetary and stellar radius, $R_p/R_\star$, estimated by the two different transit models considered. Blue points refer to the new combined model, which captures both stellar and transit signals simultaneously. The orange points refer to a simple transit model with white noise. Data points represent the relative deviation with respect to the input values and their respective uncertainties. The blue and orange solid lines and shaded areas respectively show the median and standard deviation of the corresponding data points. The black dotted line is the zero offset. Stars in the x-axis are ordered by decreasing SDE (ranging from 6.47 to 4.72).}
	\label{fig:transit_comparison}
\end{figure}

Figure~\ref{fig:transit_comparison} shows a comparison of the estimated ratio between the planetary and stellar radius, $R_p/R_\star$, when considering the two different models for the light curve. The blue solid line shows that the new combined model leads to a $-$0.06\% offset (5.80\% scatter) between the input values and the estimated ones, whilst the orange solid line denotes a 1.71\% offset (6.49\% scatter) for the simple transit model with white noise. This preliminary result suggests that the new combined model is capable of recovering both more precise and accurate planetary radii.

\section{Conclusions}\label{sec:conclusions}

Gaussian Processes were employed to model the stellar signals (i.e., granulation and oscillations) of low-luminosity red-giant stars in the time domain. Two models were considered: Model 1 contains a mesogranulation component, an oscillation bump component and a white noise component; Model 2 results from adding an extra granulation component to Model 1. Both models were applied to TESS simulated data (generated considering the presence of both mesogranulation and granulation components) to test the validity of the method and its applicability to TESS light curves. Furthermore, the models were also applied to a sample of \textit{Kepler} stars in order to compare this method with the commonly used power-spectrum fitting method.

Following the analysis and discussion of the results presented in Sections \ref{sec:appTESS} and \ref{sec:appKepler}, some conclusions can be drawn:
\begin{itemize}
	\item Application to TESS simulated data showed that GP regression is capable of capturing the stellar background signal and oscillations directly in the time domain. Due to the high white noise levels and the short duration of typical TESS time series (27.4 days), the simpler Model 1 seems to be better suited to find stellar signals. In particular, an accurate determination of $\nu_\text{max}$ was made possible, with a small bias of $\approx$1\%.
	\item Comparison with the power-spectrum fitting method using \textit{Kepler} data showed that both methods find the same stellar signals, with any disagreement in specific parameters being attributed to differences in the models adopted to capture the signal of the oscillation bump.
	\item Provided a physically motivated model is chosen, Gaussian Processes can be used to model stellar signals in the time domain, hence becoming a valid alternative to the commonly used power-spectrum fitting method.
\end{itemize}

Moreover, preliminary results from modelling injected giant-planet transits using the method presented here show that the ratio between the planetary and stellar radius, $R_p/R_\star$, can be estimated with both higher precision and accuracy, compared to a simple transit model plus white noise. Overall, these results suggest that the method described in this work provides a chance to improve on previous methodologies \citep[e.g.][]{barclay_2015,grunblatt_2017} for modelling the light curves of planetary transits together with the host star's signals by describing a more complete model of the stellar signals with physically motivated parameters. Future work will involve testing a more systematic application of the method to simulated data to assess the improvements in the estimation of all transit properties.

Additionally, the method also ties in with one of the "Important scientific opportunities for \textit{Kepler} \& K2 Data" \citep{barentsen_2018_white_paper}, specifically, on performing asteroseismology in the time domain.

The implementation of the method detailed in this article is publicly available\footnote{\url{https://github.com/Fill4/gptransits}}.

\section*{Acknowledgements}

The authors would like to thank M\'ario Jo\~{a}o Monteiro, Benard Nsamba, Mathieu Vrard and Tim Bedding for fruitful discussions. F.P.~acknowledges support from fellowship PD/BD/135227/2017 funded by FCT (Portugal) and POPH/FSE (EC). The project leading to this publication has received funding from the European Union’s Horizon 2020 research and innovation programme under the Marie Sk\l{}odowska-Curie grant agreement No.~792848 (PULSATION). T.L.C.~acknowledges support from grant CIAAUP-12/2018-BPD. M.C.~acknowledges support in the form of work contract funded by national funds through FCT - Fundação para a Ciência e Tecnologia. J.P.F.~is supported in the form of a work contract funded by national funds through FCT with reference DL57/2016/CP1364/CT0005. S.C.C.B.~acknowledges support from FCT through Investigador FCT contract IF/01312/2014/CP1215/CT0004. E.C. is funded by the European Union’s Horizon 2020 research and innovation program under the Marie Sklodowska-Curie grant agreement No. 664931. This work was supported by FCT/MCTES through national funds (PIDDAC) by this grant UID/FIS/04434/2019 and by FCT - Fundação para a Ciência e a Tecnologia through national funds and by FEDER through COMPETE2020 - Programa Operacional Competitividade e Internacionalização by these grants: PTDC/FIS-AST/30389/2017 \& POCI-01-0145-FEDER-030389; PTDC/FIS-AST/32113/2017 \& POCI-01-0145-FEDER-032113 and PTDC/FIS-AST/28953/2017 \& POCI-01-0145-FEDER-028953. The research leading to the presented results has received funding from the European Research Council under the European Community’s Seventh Framework Programme (FP7/2007-2013) / ERC grant agreement no. 338251 (StellarAges). This research was supported in part by the National Science Foundation under Grant No.~NSF PHY-1748958 through the Kavli Institute for Theoretical Physics program ``Better Stars, Better Planets''.




\bibliographystyle{mnras}
\bibliography{main}

\begin{thebibliography}{}
\makeatletter
\relax
\def\mn@urlcharsother{\let\do\@makeother \do\$\do\&\do\#\do\^\do\_\do\%\do\~}
\def\mn@doi{\begingroup\mn@urlcharsother \@ifnextchar [ {\mn@doi@}
  {\mn@doi@[]}}
\def\mn@doi@[#1]#2{\def\@tempa{#1}\ifx\@tempa\@empty \href
  {http://dx.doi.org/#2} {doi:#2}\else \href {http://dx.doi.org/#2} {#1}\fi
  \endgroup}
\def\mn@eprint#1#2{\mn@eprint@#1:#2::\@nil}
\def\mn@eprint@arXiv#1{\href {http://arxiv.org/abs/#1} {{\tt arXiv:#1}}}
\def\mn@eprint@dblp#1{\href {http://dblp.uni-trier.de/rec/bibtex/#1.xml}
  {dblp:#1}}
\def\mn@eprint@#1:#2:#3:#4\@nil{\def\@tempa {#1}\def\@tempb {#2}\def\@tempc
  {#3}\ifx \@tempc \@empty \let \@tempc \@tempb \let \@tempb \@tempa \fi \ifx
  \@tempb \@empty \def\@tempb {arXiv}\fi \@ifundefined
  {mn@eprint@\@tempb}{\@tempb:\@tempc}{\expandafter \expandafter \csname
  mn@eprint@\@tempb\endcsname \expandafter{\@tempc}}}

\bibitem[\protect\citeauthoryear{{Ambikasaran}, {Foreman-Mackey}, {Greengard},
  {Hogg}  \& {O'Neil}}{{Ambikasaran} et~al.}{2015}]{ambikasaran_2015}
{Ambikasaran} S.,  {Foreman-Mackey} D.,  {Greengard} L.,  {Hogg} D.~W.,
  {O'Neil} M.,  2015, \mn@doi [IEEE Transactions on Pattern Analysis and
  Machine Intelligence] {10.1109/TPAMI.2015.2448083}, \href
  {http://adsabs.harvard.edu/abs/2015ITPAM..38..252A} {38}

\bibitem[\protect\citeauthoryear{{Barclay}, {Endl}, {Huber}, {Foreman-Mackey},
  {Cochran}, {MacQueen}, {Rowe}  \& {Quintana}}{{Barclay}
  et~al.}{2015}]{barclay_2015}
{Barclay} T.,  {Endl} M.,  {Huber} D.,  {Foreman-Mackey} D.,  {Cochran} W.~D.,
  {MacQueen} P.~J.,  {Rowe} J.~F.,   {Quintana} E.~V.,  2015, \mn@doi [\apj]
  {10.1088/0004-637X/800/1/46}, \href
  {http://adsabs.harvard.edu/abs/2015ApJ...800...46B} {800, 46}

\bibitem[\protect\citeauthoryear{{Barentsen}, {Hedges}, {Saunders}, {Cody},
  {Gully-Santiago}, {Bryson}  \& {Dotson}}{{Barentsen}
  et~al.}{2018}]{barentsen_2018_white_paper}
{Barentsen} G.,  {Hedges} C.,  {Saunders} N.,  {Cody} A.~M.,  {Gully-Santiago}
  M.,  {Bryson} S.,   {Dotson} J.~L.,  2018, preprint, \href
  {https://ui.adsabs.harvard.edu/#abs/2018arXiv181012554B} {p.
  arXiv:1810.12554} (\mn@eprint {arXiv} {1810.12554})

\bibitem[\protect\citeauthoryear{{Brewer} \& {Stello}}{{Brewer} \&
  {Stello}}{2009}]{brewer_2009}
{Brewer} B.~J.,  {Stello} D.,  2009, \mn@doi [\mnras]
  {10.1111/j.1365-2966.2009.14679.x}, \href
  {http://adsabs.harvard.edu/abs/2009MNRAS.395.2226B} {395, 2226}

\bibitem[\protect\citeauthoryear{Brooks \& Gelman}{Brooks \&
  Gelman}{1998}]{gelman_1998}
Brooks S.~P.,  Gelman A.,  1998, \mn@doi [Journal of Computational and
  Graphical Statistics] {10.1080/10618600.1998.10474787}, 7, 434

\bibitem[\protect\citeauthoryear{{Campante} et~al.,}{{Campante}
  et~al.}{2016}]{campante_2016b}
{Campante} T.~L.,  et~al., 2016, \mn@doi [\apj] {10.3847/0004-637X/830/2/138},
  \href {http://adsabs.harvard.edu/abs/2016ApJ...830..138C} {830, 138}

\bibitem[\protect\citeauthoryear{{Campante}, {Barros}, {Demangeon}, {da
  N{\'o}brega}, {Kuszlewicz}, {Pereira}, {Chaplin}  \& {Huber}}{{Campante}
  et~al.}{2018}]{campante_2018}
{Campante} T.~L.,  {Barros} S. C.~C.,  {Demangeon} O.,  {da N{\'o}brega} H.~J.,
   {Kuszlewicz} J.~S.,  {Pereira} F.,  {Chaplin} W.~J.,   {Huber} D.,  2018,
  arXiv e-prints, \href {https://ui.adsabs.harvard.edu/abs/2018arXiv181206150C}
  {p. arXiv:1812.06150}

\bibitem[\protect\citeauthoryear{{Carter} \& {Winn}}{{Carter} \&
  {Winn}}{2009}]{carter_2009}
{Carter} J.~A.,  {Winn} J.~N.,  2009, \mn@doi [\apj]
  {10.1088/0004-637X/704/1/51}, \href
  {http://adsabs.harvard.edu/abs/2009ApJ...704...51C} {704, 51}

\bibitem[\protect\citeauthoryear{{Corsaro} \& {De Ridder}}{{Corsaro} \& {De
  Ridder}}{2014}]{corsaro_2014}
{Corsaro} E.,  {De Ridder} J.,  2014, \mn@doi [\aap]
  {10.1051/0004-6361/201424181}, \href
  {http://adsabs.harvard.edu/abs/2014A%26A...571A..71C} {571, A71}

\bibitem[\protect\citeauthoryear{{Corsaro}, {De Ridder}  \&
  {Garc{\'{\i}}a}}{{Corsaro} et~al.}{2015}]{corsaro_2015b}
{Corsaro} E.,  {De Ridder} J.,   {Garc{\'{\i}}a} R.~A.,  2015, \mn@doi [\aap]
  {10.1051/0004-6361/201525895}, \href
  {http://adsabs.harvard.edu/abs/2015A%26A...579A..83C} {579, A83}

\bibitem[\protect\citeauthoryear{{Faria}, {Haywood}, {Brewer}, {Figueira},
  {Oshagh}, {Santerne}  \& {Santos}}{{Faria} et~al.}{2016}]{faria_2016a}
{Faria} J.~P.,  {Haywood} R.~D.,  {Brewer} B.~J.,  {Figueira} P.,  {Oshagh} M.,
   {Santerne} A.,   {Santos} N.~C.,  2016, \mn@doi [\aap]
  {10.1051/0004-6361/201527899}, \href
  {https://ui.adsabs.harvard.edu/#abs/2016A&A...588A..31F} {588, A31}

\bibitem[\protect\citeauthoryear{{Farr} et~al.,}{{Farr}
  et~al.}{2018}]{farr_2018}
{Farr} W.~M.,  et~al., 2018, \mn@doi [\apj] {10.3847/2041-8213/aadfde}, \href
  {https://ui.adsabs.harvard.edu/#abs/2018ApJ...865L..20F} {865, L20}

\bibitem[\protect\citeauthoryear{{Foreman-Mackey}, {Agol}, {Ambikasaran}  \&
  {Angus}}{{Foreman-Mackey} et~al.}{2017}]{foreman-mackey_2017}
{Foreman-Mackey} D.,  {Agol} E.,  {Ambikasaran} S.,   {Angus} R.,  2017,
  \mn@doi [\aj] {10.3847/1538-3881/aa9332}, \href
  {http://adsabs.harvard.edu/abs/2017AJ....154..220F} {154, 220}

\bibitem[\protect\citeauthoryear{{Geweke}}{{Geweke}}{1992}]{geweke_1992}
{Geweke} J.,  1992, in Bayesian Statistics 4. Oxford: Oxford University Press,
  pp 169--193

\bibitem[\protect\citeauthoryear{{Gruberbauer}, {Kallinger}, {Weiss}  \&
  {Guenther}}{{Gruberbauer} et~al.}{2009}]{gruberbauer_2009}
{Gruberbauer} M.,  {Kallinger} T.,  {Weiss} W.~W.,   {Guenther} D.~B.,  2009,
  \mn@doi [\aap] {10.1051/0004-6361/200811203}, \href
  {https://ui.adsabs.harvard.edu/abs/2009A&A...506.1043G} {506, 1043}

\bibitem[\protect\citeauthoryear{{Grunblatt} et~al.,}{{Grunblatt}
  et~al.}{2016}]{grunblatt_2016}
{Grunblatt} S.~K.,  et~al., 2016, \mn@doi [\aj] {10.3847/0004-6256/152/6/185},
  \href {http://adsabs.harvard.edu/abs/2016AJ....152..185G} {152, 185}

\bibitem[\protect\citeauthoryear{{Grunblatt} et~al.,}{{Grunblatt}
  et~al.}{2017}]{grunblatt_2017}
{Grunblatt} S.~K.,  et~al., 2017, \mn@doi [\aj] {10.3847/1538-3881/aa932d},
  \href {http://adsabs.harvard.edu/abs/2017AJ....154..254G} {154, 254}

\bibitem[\protect\citeauthoryear{{Harvey}}{{Harvey}}{1985}]{harvey_1985}
{Harvey} J.,  1985, in {Rolfe} E.,  {Battrick} B.,  eds,  ESA Special
  Publication Vol. 235, Future Missions in Solar, Heliospheric \& Space Plasma
  Physics.

\bibitem[\protect\citeauthoryear{{Haywood} et~al.,}{{Haywood}
  et~al.}{2014}]{haywood_2014}
{Haywood} R.~D.,  et~al., 2014, \mn@doi [\mnras] {10.1093/mnras/stu1320}, \href
  {https://ui.adsabs.harvard.edu/abs/2014MNRAS.443.2517H} {443, 2517}

\bibitem[\protect\citeauthoryear{{Huang} et~al.,}{{Huang}
  et~al.}{2018}]{huang_2018}
{Huang} C.~X.,  et~al., 2018, preprint, \href
  {http://adsabs.harvard.edu/abs/2018arXiv180905967H} {} (\mn@eprint {arXiv}
  {1809.05967})

\bibitem[\protect\citeauthoryear{{Huber} et~al.,}{{Huber}
  et~al.}{2019}]{huber_2019}
{Huber} D.,  et~al., 2019, \mn@doi [\aj] {10.3847/1538-3881/ab1488}, \href
  {https://ui.adsabs.harvard.edu/abs/2019AJ....157..245H} {157, 245}

\bibitem[\protect\citeauthoryear{{Kallinger} et~al.,}{{Kallinger}
  et~al.}{2010}]{kallinger_2010}
{Kallinger} T.,  et~al., 2010, \mn@doi [\aap] {10.1051/0004-6361/200811437},
  \href {https://ui.adsabs.harvard.edu/abs/2010A&A...509A..77K} {509, A77}

\bibitem[\protect\citeauthoryear{{Kallinger} et~al.,}{{Kallinger}
  et~al.}{2014}]{kallinger_2014}
{Kallinger} T.,  et~al., 2014, \mn@doi [\aap] {10.1051/0004-6361/201424313},
  \href {http://adsabs.harvard.edu/abs/2014A%26A...570A..41K} {570, A41}

\bibitem[\protect\citeauthoryear{{Kallinger}, {Hekker}, {Garcia}, {Huber}  \&
  {Matthews}}{{Kallinger} et~al.}{2016}]{kallinger_2016}
{Kallinger} T.,  {Hekker} S.,  {Garcia} R.~A.,  {Huber} D.,   {Matthews} J.~M.,
   2016, \mn@doi [Science Advances] {10.1126/sciadv.1500654}, \href
  {http://adsabs.harvard.edu/abs/2016SciA....2E0654K} {2, 1500654}

\bibitem[\protect\citeauthoryear{{Kov{\'a}cs}, {Zucker}  \&
  {Mazeh}}{{Kov{\'a}cs} et~al.}{2002}]{kovacs_2002}
{Kov{\'a}cs} G.,  {Zucker} S.,   {Mazeh} T.,  2002, \mn@doi [\aap]
  {10.1051/0004-6361:20020802}, \href
  {http://adsabs.harvard.edu/abs/2002A%26A...391..369K} {391, 369}

\bibitem[\protect\citeauthoryear{{Kreidberg}}{{Kreidberg}}{2015}]{kreidberg_2015}
{Kreidberg} L.,  2015, \mn@doi [\pasp] {10.1086/683602}, \href
  {http://adsabs.harvard.edu/abs/2015PASP..127.1161K} {127, 1161}

\bibitem[\protect\citeauthoryear{{Kuszlewicz}, {Chaplin}, {North}, {Farr},
  {Bell}, {Davies}, {Campante}  \& {Hekker}}{{Kuszlewicz}
  et~al.}{2019}]{kuszlewicz_2019}
{Kuszlewicz} J.~S.,  {Chaplin} W.~J.,  {North} T. S.~H.,  {Farr} W.~M.,  {Bell}
  K.~J.,  {Davies} G.~R.,  {Campante} T.~L.,   {Hekker} S.,  2019, \mn@doi
  [\mnras] {10.1093/mnras/stz1689}, \href
  {https://ui.adsabs.harvard.edu/abs/2019MNRAS.488..572K} {488, 572}

\bibitem[\protect\citeauthoryear{{Mathur} et~al.,}{{Mathur}
  et~al.}{2011}]{mathur_2011}
{Mathur} S.,  et~al., 2011, \mn@doi [\apj] {10.1088/0004-637X/741/2/119}, \href
  {http://adsabs.harvard.edu/abs/2011ApJ...741..119M} {741, 119}

\bibitem[\protect\citeauthoryear{{Mosser} et~al.,}{{Mosser}
  et~al.}{2012}]{mosser_2012a}
{Mosser} B.,  et~al., 2012, \mn@doi [\aap] {10.1051/0004-6361/201117352}, \href
  {https://ui.adsabs.harvard.edu/abs/2012A&A...537A..30M} {537, A30}

\bibitem[\protect\citeauthoryear{{North} et~al.,}{{North}
  et~al.}{2017}]{north_2017}
{North} T.~S.~H.,  et~al., 2017, \mn@doi [\mnras] {10.1093/mnras/stw2782},
  \href {http://adsabs.harvard.edu/abs/2017MNRAS.465.1308N} {465, 1308}

\bibitem[\protect\citeauthoryear{Rasmussen \& Williams}{Rasmussen \&
  Williams}{2006}]{rasmussen_2006}
Rasmussen C.,  Williams C.,  2006, Gaussian Processes for Machine Learning.
Adaptive Computation and Machine Learning, MIT Press, Cambridge, MA, USA

\bibitem[\protect\citeauthoryear{{Ricker} et~al.,}{{Ricker}
  et~al.}{2015}]{ricker_2015}
{Ricker} G.~R.,  et~al., 2015, \mn@doi [Journal of Astronomical Telescopes,
  Instruments, and Systems] {10.1117/1.JATIS.1.1.014003}, \href
  {http://adsabs.harvard.edu/abs/2015JATIS...1a4003R} {1, 014003}

\bibitem[\protect\citeauthoryear{{Sullivan} et~al.,}{{Sullivan}
  et~al.}{2015}]{sullivan_2015}
{Sullivan} P.~W.,  et~al., 2015, \mn@doi [\apj] {10.1088/0004-637X/809/1/77},
  \href {http://adsabs.harvard.edu/abs/2015ApJ...809...77S} {809, 77}

\bibitem[\protect\citeauthoryear{{Vanderspek} et~al.,}{{Vanderspek}
  et~al.}{2018}]{vanderspek_2018}
{Vanderspek} R.,  et~al., 2018, preprint, \href
  {http://adsabs.harvard.edu/abs/2018arXiv180907242V} {} (\mn@eprint {arXiv}
  {1809.07242})

\makeatother
\end{thebibliography}



\appendix

\section{Parseval Normalization of the {\sc celerite} PSD}\label{append:normal}

The PSD in Eq.~(\ref{eq_psd_gran}),
\begin{equation}
	S(\omega) = \sqrt{\frac{2}{\pi}} \frac{S_0}{\left( \omega / \omega_0 \right)^4 + 1} \, ,
	\label{psd_celerite}
\end{equation}
shares the functional form of the PSD describing the granulation in \citet{kallinger_2014},
\begin{equation}
	S(\nu) = \frac{2\sqrt{2}}{\pi} \frac{a^2/b}{\left( \nu/b \right)^4 +1}.
	\label{psd_kallinger}
\end{equation}
However, unlike Eq.~\ref{psd_kallinger}, Eq.~\ref{psd_celerite} is not normalized according to Parseval's theorem.

For a Parseval-normalized PSD, the variance of the light curve must equal $a^2$. In order to normalize Eq.~\ref{psd_celerite}, a constant $K$ needs to be found that ensures that the previous condition is met. From \citet{foreman-mackey_2017}, the variance of a light curve described by the kernel in Eq.~(\ref{eq_kernel_gran}) is
\begin{equation}
	k(\tau = 0) = \frac{S_0\omega_0}{\sqrt{2}},
\end{equation}
which gives
\begin{equation}
	a^2 = \frac{S_0\omega_0}{\sqrt{2}}.
	\label{variance_2}
\end{equation}
Moreover, from \citet{foreman-mackey_2017}, $\omega_0$ is expressed as
\begin{equation}
	\omega_0 = 2 \pi b \, .
	\label{omega}
\end{equation}

Equating Eqs.~\ref{psd_celerite} and \ref{psd_kallinger},
\begin{equation}
	K \sqrt{\frac{2}{\pi}} \frac{S_0}{\left( \omega / \omega_0 \right)^4 + 1} = \frac{2\sqrt{2}}{\pi} \frac{a^2/b}{\left( \nu/b \right)^4 +1} \, ,
	\label{equality}
\end{equation}
and substituting for Eqs.~\ref{variance_2} and \ref{omega}, $K$ becomes
\begin{equation}
	K = 2 \sqrt{2\pi}.
\end{equation}
Having obtained a value for $K$, Eq.~\ref{psd_celerite} can be normalized according to Parseval's theorem as
\begin{equation}
	S(\omega) = \frac{4S_0}{\left( \omega / \omega_0 \right)^4 + 1} \, .
\end{equation}


\bsp	
\label{lastpage}
\end{document}